\documentclass[prb,aps,12pt]{revtex4}
\newcommand{\be}{\begin{eqnarray}}
\newcommand{\bea}{\begin{eqnarray}}
\newcommand{\ee}{\end{eqnarray}}
\newcommand{\eea}{\end{eqnarray}}

\begin{document}
\title{Universal Longtime Dynamics in Dense Simple Fluids}
\author{Gene F. Mazenko}
\affiliation{The James Franck Institute and the Department of Physics\\
The University of Chicago\\
Chicago, Illinois 60637, USA}
%date{\today}
\maketitle

%\begin{abstract}
\centerline{Abstract}

There appears to be a longtime, very slowly evolving state in dense simple
fluids which, for high enough density, approaches a glassy nonergodic state.
The nature of the nonergodic state can be characterized by the associated
static equilibrium state.  In particular, systems driven by Smoluchowski or
Newtonian dynamics share the same static equilibrium and nonergodic states.
That these systems share the same nonergodic states is a highly nontrivial
statement and requires establishing a number of results. In the
high-density regime one finds that an equilibrating system decays
via a three-step process identified in mode-coupling theory (MCT).  For densities
greater than
a critical density, $\eta^{*}$, one has  time-power-law decay with
exponents $a$ and $b$.
A key ingredient in identifying the universal nature of the long-time results
is the recognition that the theory can be expressed in terms of two
fundamental fields: the particle density $\rho$ and the response field $B$.
There are sets of linear fluctuation dissipation relations (FDRs) which connect
the cumulants of these two fields.  The form of the FDRs is the same for
both Smoluchowski or Newtonian dynamics.
While we show this universality of nonergodic states
within perturbation theory,
we expect it to be true more generally.

The nature of the approach to the nonergodic state has been suggested by MCT.
It has been a point of contention that MCT is a phenomenological theory
and not a systematic theory with prospects for
improvement.  Recently a systematic  theory has been developed.  It naturally
allows one to calculate self-consistently
density cumulants in a perturbation expansion in a
pseudo-potential.
At leading order one obtains a kinetic kernel quadratic in the density.
This is a ``one-loop" theory like MCT.  At this one-loop level one finds vertex
corrections which depend on the three-point equilibrium cumulants.
Here we assume these vertex-corrections can be ignored. We focus on the higher-
order loops here.  We show that one can sum up all of the loop contributions.
The higher-order loops do not change the nonergodic state parameters substantially.

%\end{abstract}

%\pacs{PACS numbers: 05.70.Ln, 64.60.Cn, 64.60.My, 64.75.+g}

\section{Introduction}

There is a large body of evidence\cite{BB} that dense simple fluids show an
interesting long-time slow dynamics.
We develop here the ingredients of a theory of
the universal features
of this long-time dynamics.
The theory is applicable to systems evolving under Newtonian\cite{NDI,NDII},
Smoluchowski\cite{FTSPD,SDENE} and, possibly,
Fokker-Planck dynamics\cite{FPD} in equilibrium.
Each of these systems
share the same thermodynamics and equilibrium static structure.
We conjecture that there is a long-time slow kinetics regime governed
by the solutions of what we call the ergodic-nonergodic problem (ENEP).

In this paper we formulate the ENEP within the self-consistent
perturbation theory introduced in FTSPD\cite{FTSPD},
SDENE\cite{SDENE}, NDI\cite{NDI}, NDII\cite{NDII}, MMSI\cite{MMSI} and
SM\cite{SM}.
We show, within perturbation theory,
that both Smoluchowski dynamics (SD) and
Newtonian dynamics (ND) lead to the same ENEP.
We explore the degree to which the long-time kinetics
of classes of simple fluids depend only on the nature of the
interaction potential and not on the details of the kinetic process.

Given the form of an ENEP one must still, for each interaction potential,
determine if the system undergoes an ENE transition and shows the
famous two-step kinetic process explored in conventional mode-coupling
theory (MCT)\cite{MCT1,MCT2,SM}.
Here we establish the form of the ergodic-nonergodic problem for simple
fluids.
In previous work in SDENE we
established the ENEP for SD for the case where the dynamical vertices
were evaluated using the noninteracting ideal gas dynamic three-point
vertices. Here we find the appropriate ENEP for both SD and ND
including vertex corrections
and determine those properties which determine the members of a universality
class.

We show that the ENEP has two sets of contributions.
First there is a backbone loop structure which includes the one-loop
contribution familiar from MCT.  Upon this loop structure rests a set
of vertex corrections.  We show here how to add up all of the loop
contributions.  Explicit evaluation of the vertex corrections will be
carried out elsewhere.
Within our systematic development here, MCT is a one-loop
theory without vertex corrections.  Note the {\it vertices}
in MCT differ from those found here.

We also show how to practically solve the ENEP in the rather
sophisticated
approximation where one treats all loops but
ignores vertex corrections. We call this the no-vertex correction (NVC)
approximation.  In this approximation we removed the constraint
that the pseudo-potential, used in perturbation theory, be small.

In working out the NVC approximation
in some completeness we have learned
about the shared kinetic structure that leads to the universal
nature of the ENEP.  It is interesting (but not surprising) that
with the proper identification of core variables, (the density
$\rho$ and a response field $B$), the matrix of  cumulants among
the variables $\rho$ and $B$,
share the same fluctuation-dissipation relations.
(The first treatment of these higher cumulants is given
in NDII.)
This is despite the fact
that the microscopic definitions of $B$ are quite different for
SD and ND.  What is much less obvious is that the three-point
cumulants and vertices for both SD and ND obey \cite{NDII}
eight new fluctuation-
dissipation relations or identities among them.  These identities
are of the same form for both sets of dynamics.  We conjecture
that one will find the same set of identities for Fokker-Planck
dynamics.  This is an exact nonperturbative set of results.
These statements are true despite the fact the ideal gas realizations
of, for example, $G^{(0)}_{B\rho\rho}$ are very different for SD and ND.

In carrying out the analysis of the self-energies (memory function)
and showing that they collapse onto a single ENEP, it is crucial
that one use these identities and certain theorems derived from them.

Roughly speaking, if one chooses the core variables for each system
properly, then cumulants for two-systems
share the same FDRs, thermodynamics,
equilibrium static structure, and, we show, the same ENEP.

\section{Class of Kinetic Systems}

We discuss two rather different particle kinetics.  For a system
of $N$ particles with mass $m$, at positions $R_{i}$ and momenta $P_{i}$
we can consider
Newtonian dynamics, with equations of motion
\be
m\dot{R}_{i}=P_{i}
\ee
and
\be
\dot{P}_{i}=f_{i}
~~~,
\ee
or
Smoluchowski dynamics, with equations of motion
\be
\dot{R}_{i}=Df_{i}+\xi_{i}
~~~.
\ee
The force in both cases is given by
\be
f_{i}=-\nabla_{i}U
~~~,
\ee
where the total potential energy is
\be
U=\frac{1}{2}\sum_{i,j}V(R_{i}-R_{j})
\ee
where $V(r)$ is a standard pair potential.
$D$ is a diffusion coefficient and $\xi_{i}$ is gaussian noise
with variance
\be
\langle \xi_{i}(t)\xi_{j}(t')\rangle =2Dk_{B}T\delta_{i,j}\delta (t-t')
~~~.
\ee
Starting with these equations of motion we can reexpress each problem in terms
of a path-integral formulation.  The SD are treated in FTSPD and ND in ND1.
Both problems
can be organised around a field-theoretic treatment of a core problem.
All the members of a class have the same equilibrium
static structure. In particular they share the same static equilibrium
density correlation functions:
\be
C_{\rho\rho\ldots\rho}(x_{1},x_{2},\ldots,x_{n})
=\langle \rho (x_{1})\rho (x_{2}),\ldots,\rho (x_{n})\rangle
~~~.
\ee

We now set up the general long-time kinetics in terms of a set of
core variables;
$\Phi =(\rho , B)$.
The core variables in these systems are the
physical observables the number density,
\be
\rho (1)=\sum_{i = 1}^{N}\delta (x_{1} - R_{i}(t_{1}))
\ee
and the crucial response field $B(1)$ which depends on the
particle dynamics, SD or ND.

In FTSPD and ND1 we introduced a field theoretical formulation of classical
many-particle dynamics.  The grand canonical partition function is of
the general form
\be
Z=\sum_{N=0}^{\infty}\frac{z^{N}}{N!}Tr e^{-A+H\cdot\Phi}
\ee
where the trace is over the phase-space degrees of freedom plus the
associated MSR\cite{MSR} conjugate fields, $A$ is the MSR action given by
\be
A = A_{0}+\frac{1}{2}\sum_{ij}\Phi_{i}\sigma_{ij}\Phi_{j}
\label{eq:10}
\ee
where $i$ and $j$ labels space, time and
fields $\rho$ or $B$,
and $A_{0}$ is the noninteracting MSR action and the interacting
contribution can be written in terms of the {\it core} variables
$\Phi$ which essentially
defines the response fields $B$.
The interaction matrix $\sigma_{ij}$ is given below.
$H_{i}$ is an external-time and space-dependent field that couples to the
core variables $\Phi_{i}$.

The generator of dynamical cumulants is given by
\be
W[H]=ln Z[H]
~~~.
\ee
The cumulants are given by
\be
G_{ij\ldots k}=
\frac{\delta \delta \ldots\delta W[H]}
{\delta H_{i}\delta H_{j}\ldots \delta H_{k}}
\label{eq:12}
~~~.
\ee
It was shown in FTSPD and NDI that the one-point average
\be
G_{i}=\langle \Phi_{i}\rangle
=\frac{\delta}{\delta H_{i}}W[H]
\ee
satisfies the fundamental identity
\be
G_{i}=\tilde{T}r \phi_{i} e^{H\cdot\phi +\Delta W [H]}
\label{eq:194}
\ee
where  $\tilde{T}r$ is the sum over all the defrees of freedom including
the conjugate MSR\cite{MSR} degrees of freedom for a typical particle.
For the core variables we write
\be
\Phi_{i}=\sum_{\alpha =1}^{N}\phi_{i}^{\alpha}
\ee
where $\alpha$ labels the particles and
\be
\phi_{\rho}^{(0)}(1)=\delta (x_{1}-R^{(0)}(t_{1}))
\ee
and $\phi_{B}^{(0)}(1)$ is strongly dependent on the type of dynamics.
For SD dynamics see FTSPD while for ND see
ND1.
Finally we have in Eq.(\ref{eq:194})
\be
\Delta W[H] =W[H+F]-W[H]
\label{eq:17}
\ee
with
\be
F_{i}=\sum_{j}\sigma_{ij}\phi_{j}
\label{eq:45}
\ee
and
where for systems fluctuating in equilibrium the force matrix,
appearing in Eq.(\ref{eq:10}), is given
in terms of fourier transforms by
\be
\sigma_{\alpha\beta}(q)=V(q)[\delta_{\alpha \rho}
\delta_{\beta B}+\delta_{\alpha B}\delta_{\beta \rho}]
~~~.
\ee

\section{Fluctuation Dissipation Relations}

\subsection{General Results}

In this section we explore the nature of
fluctuation-dissipation relations (FDR)
obeyed by our class of systems. In ND2 we find the general FDR
for Newtonian dynamics for the core fields
$\rho (x,t)$ and  $B(x,t)$.
We found the simple result for n-point cumulants in terms of Fourier
transforms:
\be
G_{BB...B\rho\rho}(12\ldots\ell,\ell+1\ldots n)
=G^{*}_{\psi\psi...\psi\rho\ldots\rho}(12\ldots\ell,\ell+1\ldots n)
\label{eq:20}
\ee
where
\be
\psi_{i}=B_{i}-i\beta\omega_{i}\rho_{i}
\ee
and for the single-particle quantities
\be
{\cal G}_{BB...B\rho\rho}(12\ldots\ell,\ell+1\ldots n)
={\cal G}^{*}_{\psi\psi...\psi\rho\ldots\rho}(12\ldots\ell,\ell+1\ldots n)
~~~.
\ee
In the case of SD these relations have been shown to hold
for cumulants with up to five $B$ insertions\cite{MMS2}.
It seems likely that they
hold in general.
Do these results also hold for Fokker-Planck dynamics?
These FDRs are extremely useful.

\subsection{Two-point cumulants}

Using the FD relations at the two-point level we find
for the two-point cumulant:
\be
G_{B\rho}(12)=G^{*}_{B\rho}(12)
+i\beta\omega_{1}G^{*}_{\rho\rho}(12)
\label{eq:3.1}
\ee
where $1=(q_{1},\omega_{1})$.
Due to translational invariance in space and time
\be
G_{\alpha\beta}(12)
=G_{\alpha\beta}(1)\delta (1+2)
\ee
and
\be
G_{B\rho}(1)
=G^{*}_{B\rho}(1)
~~~.
\label{eq:25}
\ee
We can also show that
\be
G_{BB}(12)=0
~~~.
\label{eq:37a}
\ee
We see that Eq.(\ref{eq:3.1}) can be written as
\be
G_{B\rho}(1)=G^{*}_{B\rho}(1)
+i\beta\omega_{1}G^{*}_{\rho\rho}(1)
~~~.
\label{eq:3.1}
\ee
Since $G_{\rho\rho}(1)$ is real, we have the
conventional result
\be
Im G_{B\rho}(1)= -Im G_{\rho B}(1)=\frac{\beta\omega_{1}}{2}G_{\rho\rho}(1)
\ee
which can be used to construct the dispersion relation
\be
G_{\rho B}(1)=\int\frac{d\bar{\omega}}{2\pi}
\frac{\beta\bar{\omega}G_{\rho\rho}(q_{1},\bar{\omega})}
{\omega_{1}-\bar{\omega}+i\eta}
~~~.
\label{eq:35}
\ee
Eq.(\ref{eq:35}) leads to
the static sum rule
\be
G_{\rho B}(q_{1},0)=-\beta S(q_{1})
\label{eq:30}
\ee
where $S(q_{1})$ is the static structure factor.
Taking the inverse time Fourier transform of Eq.(\ref{eq:35}) we find
one of the standard FDRs:
\be
G_{\rho B}(q,t)=\theta (t)\beta\frac{\partial}{\partial t}
G_{\rho\rho}(q,t)
~~~.
\ee
Combining Eqs.(\ref{eq:3.1}) and (\ref{eq:25}) we can write the
very useful result
\be
+i\beta \omega G_{\rho\rho}(q,\omega )
=G_{B\rho}(q,\omega )-G_{\rho B}(q,\omega )
\label{eq:32}
~~~.
\ee
The matrix propagator, $G_{\alpha \beta}$, is the first of a number
of two-point FDR matrix propagators (FDRMP) defined in Appendix B.
In addition to $G_{\alpha \beta}$ we have
${\cal G}_{\alpha \beta}$,
$\bar{G}_{\alpha \beta}$,
$\tilde{G}_{\alpha \beta}$,  and
noninteracting counterparts enter our analysis.

\subsection{Two-point Vertices}

The two-point vertices $\Gamma_{\alpha\mu}(1)$ are defined by
Dyson's equation:
\be
\sum_{\mu}\Gamma_{\alpha\mu}(1)G_{\mu \beta}(1)=\delta_{\alpha\beta}
~~~.
\ee
This matrix equation reduces, using Eq.(\ref{eq:37a}), to
\be
\Gamma_{\rho\rho}(1)=0
\ee
\be
\Gamma_{\rho B}(1)G_{B\rho}(1)=1
\label{eq:42}
\ee
\be
\Gamma_{B\rho }(1)G_{\rho B}(1)=1
\label{eq:43}
\ee
and
\be
\Gamma_{B B}(1)G_{B\rho}(1)
+\Gamma_{B \rho}(1)G_{\rho\rho}(1)=0
~~~.
\label{eq:44}
\ee
Eq.(\ref{eq:44}) can be written in the form
\be
G_{\rho\rho}(1)=-G_{\rho B}(1)
\Gamma_{B B}(1)G_{B\rho}(1)
~~~.
\label{eq:45}
\ee
We easily find
\be
\Gamma_{B\rho}(q,\omega )
=\Gamma^{*}_{\rho B}(q,\omega )
~~~.
\ee
Starting with Eq.(\ref{eq:3.1}),
we can use Eqs.(\ref{eq:42}) and (\ref{eq:43}) to write
\be
\frac{1}{\Gamma_{\rho B}(1)}=
\frac{1}{\Gamma_{B\rho }(1)}
+i\beta \omega_{1}G_{\rho\rho}(1)
\label{eq:118}
~~~.
\ee
Using Eq.(\ref{eq:45})
and
cancelling a common denominator,
gives
\be
i\beta\omega_{1}\Gamma_{BB}(1)
=\Gamma_{\rho B}(1)
-\Gamma_{B\rho}(1)
\ee
which can be compared with Eq.(\ref{eq:32}).

\subsection{FD Relations and Three-point cumulants}

There are substantial differences in specific cumulants for the
different realizations of simple fluid dynamics. For example
the noninteracting
three-point cumulant $G^{(0)}_{B\rho\rho}(123)$ for SD and ND
are quite different.

The FD relations,  Eq.(\ref{eq:20}), for three-point cumulants,
reduce to eight independent identities:
\be
G_{BBB}=G_{BBB}^{*}+i\beta\omega_{1}G_{\rho BB}^{*}
+i\beta\omega_{2}G_{B\rho B}^{*}
+i\beta\omega_{3}G_{BB\rho }^{*}
\label{eq:42}
\ee
\be
-\beta^{2}\left(\omega_{1}\omega_{2}G_{\rho\rho B}^{*}
+\omega_{1}\omega_{3}G_{\rho B\rho }^{*}
+\omega_{2}\omega_{3}G_{B\rho \rho }^{*}\right)
\nonumber
-i\beta^{3}\omega_{1}\omega_{2}\omega_{3}G_{\rho\rho \rho }^{*}
\ee
\be
G_{BB\rho }=G_{BB\rho }^{*}+i\beta\omega_{1}G_{\rho B\rho }^{*}
+i\beta\omega_{2}G_{B\rho \rho }^{*}
-\beta^{2}\omega_{1}\omega_{2}G_{\rho \rho \rho }^{*}
\label{eq:BBR}
\ee
\be
G_{B\rho B }=G_{B\rho B}^{*}+i\beta\omega_{1}G_{\rho \rho B}^{*}
+i\beta\omega_{3}G_{B\rho \rho }^{*}
-\beta^{2}\omega_{1}\omega_{3}G_{\rho \rho \rho }^{*}
\label{eq:BRB}
\ee
\be
G_{\rho B B }=G_{\rho BB}^{*}+i\beta\omega_{2}G_{\rho \rho B}^{*}
+i\beta\omega_{3}G_{\rho B\rho }^{*}
-\beta^{2}\omega_{2}\omega_{3}G_{\rho \rho \rho }^{*}
\label{eq:RBB}
\ee
\be
G_{B\rho \rho }=G_{B\rho \rho}^{*}+i\beta\omega_{1}G_{\rho \rho \rho}^{*}
\ee
\be
G_{\rho B\rho }=G_{\rho B\rho}^{*}+i\beta\omega_{2}G_{\rho \rho \rho}^{*}
\ee
\be
G_{\rho \rho B}=G_{\rho \rho B}^{*}+i\beta\omega_{3}G_{\rho \rho \rho}^{*}
\ee
Each of these constraints give an identity relating the
3-point cumulants.
Since we know $G_{BBB}=G_{BBB}^{*}=0$,
Eq.(\ref{eq:42}) reduces to
\be
0= i\beta\omega_{1}G_{\rho BB}^{*}
+i\beta\omega_{2}G_{B\rho B}^{*}
+i\beta\omega_{3}G_{BB\rho }^{*}
\nonumber
\ee
\be
-\beta^{2}\left(\omega_{1}\omega_{2}G_{\rho\rho B}^{*}
+\omega_{1}\omega_{3}G_{\rho B\rho }^{*}
+\omega_{2}\omega_{3}G_{B\rho \rho }^{*}\right)
\nonumber
-i\beta^{3}\omega_{1}\omega_{2}\omega_{3}G_{\rho\rho\rho}
~~~.
\label{eq:BBB}
\ee

As shown in ND2,
these cumulant identities lead to useful relations satisfied by the
three-point vertices.
The relationships connecting the three-point cumulants
and the three-point vertex functions are given by
\be
\Gamma_{\alpha\mu\nu}(123)=-\sum_{\alpha'\mu'\nu'}
\Gamma_{\alpha\alpha'}(1)\Gamma_{\mu\mu'}(2)\Gamma_{\nu\nu'}(3)
G_{\alpha'\mu'\nu'}(123)
\ee
If we introduce some simplfying notation:
\be
\phi_{1}=i\beta \omega_{1}\Gamma_{B\rho\rho}
\ee
\be
\phi_{2}=i\beta \omega_{2}\Gamma_{\rho B\rho}
\ee
\be
\phi_{3}=i\beta \omega_{3}\Gamma_{\rho \rho B}
\ee
\be
\psi_{1}=\beta^{2} \omega_{2}\omega_{3}\Gamma_{\rho BB}
\ee
\be
\psi_{2}=\beta^{2} \omega_{1}\omega_{3}\Gamma_{B\rho B}
\ee
\be
\psi_{3}=\beta^{2} \omega_{1}\omega_{2}\Gamma_{BB\rho }
\ee
\be
P=
-i\beta^{3}\omega_{1}\omega_{2}\omega_{3}\Gamma_{BBB}
~~~,
\ee
then we find, after  algebra, the FD vertex relations:
\be
P+\psi_{3}+\psi_{2}+\psi_{1}
+\phi_{3}+\phi_{2}+\phi_{1}=0
\ee
\be
-\phi_{3}^{*}+\phi_{1}+\phi_{2}+\psi_{3}=0
\ee
\be
-\phi_{2}^{*}+\phi_{1}+\phi_{3}+\psi_{2}=0
\ee
\be
-\phi_{1}^{*}+\phi_{2}+\phi_{3}+\psi_{1}=0
~~~.
\ee

\section{Kinetic Equation}

Now we start working toward a description we can use to identify the
ENE problem.  As a first step
we want to obtain the form of
$\Gamma_{\alpha\beta}(q,\omega )$
in the time domain. If we look at
\be
\Gamma_{B\rho}(q, \omega)=\frac{1}{G_{\rho B}(q,\omega)}
\label{eq:61}
\ee
and note that the response functions vanish algebraically
as $\omega\rightarrow\infty$, we see that the two-point vertex
diverges in this limit.  We assume that
\be
\lim_{\omega\rightarrow\infty}
\Gamma_{B\rho}(q, \omega)=-D_{q}\omega^{2}+i\omega A_{q}+C_{q}
\ee
where the coefficients $A$, $C$ and $D$ are real.
Define  the "local" quantity
\be
\Gamma^{(\ell )}_{B\rho}(q, \omega)=-D_{q}\omega^{2}+i\omega A_{q}+C_{q}
\label{eq:153}
\ee
for all frequencies. We also define the subtracted quantities:
\be
\Gamma^{(s )}_{B\rho}(q, \omega)=
\Gamma_{B\rho}(q, \omega)-
\Gamma^{(\ell )}_{B\rho}(q, \omega)
\ee
which vanishes for large frquencies.  At low frequencies
\be
\Gamma^{(s )}_{B\rho}(q, 0)=
\Gamma_{B\rho}(q, 0)-
\Gamma^{(\ell )}_{B\rho}(q, 0 )
\ee
where, using Eqs.(\ref{eq:30}) and (\ref{eq:61})
\be
\Gamma_{B\rho}(q, 0)=-(\beta S(q))^{-1}
\ee
and
\be
\Gamma^{(\ell )}_{B\rho}(q, 0 )=C_{q}.
\ee
This leads to
\be
\Gamma^{(s )}_{B\rho}(q, 0)=
-(\beta S(q))^{-1}-C_{q}
~~~.
\ee
We assume that the FDR holds for $\Gamma_{\alpha\beta}^{(\ell )}$ and
\be
\beta\omega \Gamma^{(\ell )}_{BB}(q, 0 )=-2\omega A_{q}
\ee
or
\be
\beta\Gamma^{(\ell )}_{BB}(q, 0 )=-2 A_{q}
\ee
and we assume $A_{q}\geq 0$.

Taking the inverse Fourier transform of Eq.(\ref{eq:153}) gives
\be
\Gamma^{(\ell )}_{B\rho}(q,t-t' )=
\left[D_{q}\frac{\partial^{2}}{\partial t ^{2}}
-A_{q}\frac{\partial}{\partial t}
+C_{q}\right]\delta (t - t')
\ee
and
\be
\beta\Gamma^{(\ell )}_{BB}(q,t-t' )=
-A_{q}\delta (t - t')
~~~.
\ee
We have then
\be
\Gamma_{B\rho}(q,t-t' )=
\Gamma^{(\ell )}_{B\rho}(q,t-t' )+
\Gamma^{(s)}_{B\rho}(q,t-t' )
\ee
\be
\Gamma_{BB}(q,t-t' )=
-\beta^{-1}A_{q}\delta (t-t' )+
\Gamma^{(s )}_{BB}(q,t-t' )
~~~.
\ee
It then follows for the subtracted parts that we have the dispersion relation
\be
\Gamma^{(s)}_{B\rho }(q,\omega )=
\int\frac{d\bar{\omega}}{2\pi}
\frac{\beta\bar{\omega}\Gamma^{(s)}_{BB}(q,\bar{\omega} )}
{\omega -\bar{\omega}+i\eta}
\ee
reflecting the fact that $\Gamma_{\alpha\beta}^{(s)}$
is analytic in the
upper-half plane and vanishes as $\omega \rightarrow \infty$.
In the time domain the FDR reads
\be
\Gamma^{(s)}_{B\rho }(q,t-t')=\beta\theta (t-t')
\frac{\partial}{\partial t}
 \Gamma^{(s)}_{BB}(q,t-t' )
~~~.
\ee

Now express Eq.(\ref{eq:44}) in the time-domain
\be
\hat{\Gamma}^{(\ell )}_{B\rho}(t)G_{\rho\rho}(t-t')
+\hat{\Gamma}^{(\ell )}_{BB}(t)G_{B\rho}(t-t')
=\Psi(t,t')
\ee
where the $q$-dependence is surpressed and
\be
\Psi (t,t')=
\int_{-\infty}^{t}ds \Sigma_{B\rho}(t-s)G_{\rho\rho}(s-t')
+\int_{-\infty}^{t'}ds
\Sigma_{BB}(t-s)G_{\rho B}(t'-s)
~~~.
\ee
The self-energies are defined by
\be
\Sigma_{\alpha\beta}(\omega )=-\Gamma_{\alpha\beta}^{(s)}(\omega)
\ee
and we have
used the fact that $\Sigma_{B\rho}(t-s)\approx \theta (t-s)$
and $G_{\rho B}(t'-s)\approx \theta (t'-s)$.  We then use the
fluctuation dissipation  relations
\be
\Sigma_{B\rho}(t-s)=\theta (t-s)\frac{\partial}{\partial t}
\beta\Sigma_{BB}(t-s)
\ee
\be
G_{\rho B}(t'-s)=\theta (t'-s)\frac{\partial}{\partial t'}
\beta G_{\rho\rho}(t'-s)
\ee
to obtain
\be
\Psi (t,t')=
-\int_{-\infty}^{t}ds \left[\frac{\partial}{\partial s}
\beta \Sigma_{BB}(t-s)\right]
G_{\rho\rho}(s-t')
\nonumber
\ee
\be
-\int_{-\infty}^{t'}ds
\Sigma_{BB}(t-s)
\frac{\partial}{\partial s}
\beta  G_{\rho\rho}(t'-s)
~~~.
\ee
Integrating by parts in the first integral:
\be
\Psi (t,t')=- \beta\Sigma_{BB}(0)G_{\rho\rho}(t-t')
+\int_{-\infty}^{t}ds\beta\Sigma_{BB}(t-s)
\frac{\partial}{\partial s}G_{\rho\rho}(s-t')
\nonumber
\ee
\be
-\int_{-\infty}^{t'}ds
\beta\Sigma_{BB}(t-s)
\frac{\partial}{\partial s}
G_{\rho\rho}(t'-s)
\nonumber
\ee
\be
=-\beta\Sigma_{BB}(0)G_{\rho\rho}(t-t')
+\int_{t'}^{t}ds \beta\Sigma_{BB}(t-s)
\frac{\partial}{\partial s}
G_{\rho\rho}(t'-s)
\ee
where we assume $t > t'$.
We then have, putting in the $q$-depence, the kinetic equation
\be
\hat{\Gamma}_{B\rho}^{(\ell)}(q,t)G_{\rho\rho}(q,t-t')
=-\beta\Sigma_{BB}(q,t=0)G_{\rho\rho}(q,t-t')
\nonumber
\ee
\be
+\int_{t'}^{t}ds \beta\Sigma_{BB}(q,t-s)
\frac{\partial}{\partial s}
G_{\rho\rho}(q,t'-s)
\label{eq:101}
\ee
where for $t > t'$ we can drop the
$\Gamma_{BB}^{(\ell )}(q)G_{B\rho}(q, t-t')$ term.
Putting all these results together we have the kinetic equation
\be
\left[D_{q}\frac{\partial^{2}}{\partial t^{2}}-A_{q}
\frac{\partial}{\partial t}-\beta^{-1}S_{q}^{-1}\right]
G_{\rho\rho}(q,t)
-\int_{t'}^{t}ds \beta\Sigma_{BB}(q,t-s)
\frac{\partial}{\partial s}
G_{\rho\rho}(q,t'-s)
=0
\label{eq:85}
~~~.
\ee
We see that our dynamical problem is now in the form of a
memory function equation and the dynamic part of the memory function
is given by the self-energy $\Sigma_{BB}$.
The static structure factor is the same for all fluids
essentially by definition.  The coeffients $A_{q}$ and $D_{q}$
can be constructed using perturbation theory. Let us look at
the two cases studied so far.

At zeroth-order the two-point
vertex is given in the SD case by
\be
\gamma_{B\rho}^{(0)}(q,\omega )=
-\frac{(-i\omega +\bar{D}q^{2})}
{\beta\bar{\rho}\bar{D}q^{2}}
~~~.
\ee
From Eq.(\ref{eq:153}) we can read-off
\be
D^{SD}_{0}(q)=0
\ee
\be
A^{SD}_{0}(q)=
\frac{1}
{\beta\bar{\rho}\bar{D}q^{2}}
\ee
and
\be
C^{SD}_{0}(q)=
-\frac{1}
{\beta\bar{\rho}}
~~~.
\ee

In the ND case we have the more complicated result:
\be
\gamma_{B\rho}^{(0)}(q,\omega )=
-\frac{1 }
{\beta\bar{\rho}\bar{S}^{*}(z)}
\ee
where
\be
\bar{S}(z)=1-2ze^{-z^{2}}\int_{0}^{z}due^{u^{2}}-i\sqrt{\pi}
ze^{-z^{2}}
\ee
and
\be
z=\frac{\omega}{\sqrt{2}qV_{0}}
\ee
with $mV_{0}^{2}=k_{B}T$. In the large frequency limit
\be
\gamma_{B\rho}^{(0)}(q,\omega )=
-\frac{1 }
{\beta\bar{\rho}}\left[-\left(\frac{\omega}{qV_{0}}\right)^{2}+1\right]
\ee
and we can identify
\be
D^{ND}_{0}(q)=
-\frac{1}
{\beta\bar{\rho}}\left(\frac{1}{qV_{0}}\right)^{2}
\ee
\be
A^{ND}_{0}(q)=0
\ee
\be
C^{ND}_{0}(q)=
-\frac{1}
{\beta\bar{\rho}}
~~~.
\ee

The kinetic equation, Eq.(\ref{eq:85}),
is diagonalized using a Laplace transform in time.  We define
\be
\hat{G}_{\rho\rho}(q,z)=\int_{0}^{\infty}dte^{-zt}G_{\rho\rho}(q, t)
\ee
and we need the results
\be
\int_{0}^{\infty}dte^{-zt}\dot{G}_{\rho\rho}(q, t)
=z\hat{G}_{\rho\rho}(q,z)-S(q)
\ee
and
\be
\int_{0}^{\infty}dte^{-zt}\ddot{G}_{\rho\rho}(q, t)
=z[z\hat{G}_{\rho\rho}(q,z)-S(q)]
~~~.
\ee
Laplace transforming the kinetic equation gives
\be
D(q)z[z\hat{G}_{\rho\rho}(q,z)-S(q)]
-A(q)[z\hat{G}_{\rho\rho}(q,z)-S(q)]
-\beta^{-1}S^{-1}(q)\hat{G}_{\rho\rho}(q,z)
\nonumber
\ee
\be
-\beta\hat{\Sigma}_{BB}(q,z)
[z\hat{G}_{\rho\rho}(q,z)-S(q)]
=0
\ee
Dividing by
$[z\hat{G}_{\rho\rho}(q,z)-S(q)]$ gives
\be
D(q)z-A(q)
-\frac{\beta^{-1}S^{-1}(q)\hat{G}_{\rho\rho}(q,z)}
{[z\hat{G}_{\rho\rho}(q,z)-S(q)]}
-\beta\hat{\Sigma}_{BB}(q,z)
=0
~~~.
\label{eq:101}
\ee
We assume in the long-time limit and for high densities that
$\hat{G}_{\rho\rho}(q,z)$ and
$\hat{\Sigma}_{BB}(q,z)$
blow up as $z\rightarrow 0$.  So the $D(q)z-A(q)$ term can be dropped.
The kinetic equation, Eq.(\ref{eq:101}),  reduces to
\be
\frac{\hat{G}_{\rho\rho}(q,z)}
{[z\hat{G}_{\rho\rho}(q,z)-S(q)]}
=-\beta^{2}S(q)\hat{\Sigma}_{BB}(q,z)
~~~.
\label{eq:171}
\ee

We must now turn to the mechanism which produces large
$\hat{G}_{\rho\rho}(q,z)$ and
$\hat{\Sigma}_{BB}(q,z)$ as $z\rightarrow 0$.
This involves determining $\hat{\Sigma}_{BB}$ as a functional
of $\hat{G}_{\rho\rho}$.
Solution of the resulting self-consistent equation, Eq.(\ref{eq:171}),
for $\hat{G}_{\rho\rho}$ corresponds to what we call the ENE problem.

\section{Determination of Self-energies}

\subsection{Dynamic Ornstein-Zernike Relation}

We now want to construct the self-energy $\Sigma_{BB}$
self-consistently as a functional of $G_{\rho\rho}$.
The first step is the
the derivation\cite{FTSPD} of a
Dynamic Ornstein-Zernike Relation.

Inserting Eq.(\ref{eq:194}) in
\be
G_{ij}=\frac{\delta}{\delta H_{j}}G_{i}
\nonumber
~~~,
\ee
we can use the chain-rule for functional differentiation to obtain
\be
G_{ij}={\cal G}_{ij}+\sum_{k}c_{ik}G_{kj}
\label{eq:47}
\ee
where the single-particle FDRMP is given by
\be
{\cal G}_{ij}=\tilde{T}r \phi_{i}\phi_{j}e^{H\cdot\phi +\Delta W}
\label{eq:54}
\ee
and
\be
c_{ij}=\tilde{T}r \phi_{i}e^{H\cdot\phi +\Delta W}
\frac{\delta}{\delta G_{j}}\Delta W.
\label{eq:34}
\ee
Since $\Delta W$ can be treated as a functional of $G_{i}$ we see
at this stage that we have available a self-consistent theory.
If we define the matrix-inverses
\be
\sum_{k}\Gamma_{ik} G_{kj}=\delta_{ij}
\label{eq:56}
\ee
and
\be
\sum_{k}\gamma_{ik}{\cal G}_{kj}=\delta_{ij}
\label{eq:39}
\ee
then the two-point vertex is given without approximation by
\be
\Gamma_{ij}=\gamma_{ij}+K_{ij}
\label{eq:50}
\ee
where
\be
K_{ij}=-\sum_{k}\gamma_{ik}c_{kj}
\label{eq:38}
\ee
is the collective contribution to the two-point vertex.

\newpage

\subsection{General Analysis}

Much of the analysis of the two-point vertex can be carried out to
all orders in an expansion in the potential.
In general the two-point vertex is given by
\be
\Gamma_{ij}=\gamma_{ij}-\sigma_{ij}-\Sigma_{ij}
\nonumber
\ee
where the collective self-energy, $\Sigma_{ij}$, is given by
Eqs.(\ref{eq:38}),
(\ref{eq:34}),
and (\ref{eq:39}),
(sums over repeated indices
implied)
\be
\Sigma_{ij}=
\gamma_{ik}\tilde{T}r \phi_{k}e^{H\cdot\phi+\Delta W}
\frac{\delta}{\delta G_{j}}
\sum_{n=2}^{\infty}\frac{1}{n!}F_{u_{1}}F_{u_{2}}\dots F_{u_{n}}
G_{u_{1}u_{2}\ldots u_{n}}
\nonumber
\ee
where we have used
Eqs.(\ref{eq:17}) and
(\ref{eq:12}) and generated the functional power-series in $F_{i}$.
Since $F_{i}$ is defined by
Eq.(\ref{eq:45}) and is independent of $G_{i}$, we have
\be
\Sigma_{ij}
=\gamma_{ik}\tilde{T}r \phi_{k}e^{H\cdot\phi+\Delta W}
\sum_{n=2}^{\infty}\frac{1}{n!}\sigma_{u_{1}v_{1}}\phi_{v_{1}}
\sigma_{u_{2}v_{2}}\phi_{v_{2}}
\dots
\sigma_{u_{n}v_{n}}\phi_{v_{n}}
\frac{\delta}{\delta G_{j}}
G_{u_{1}u_{2}\ldots u_{n}}
~~~.
\ee
After defining the single-particle quantity
\be
{\cal G}_{kv_{1}v_{2}\ldots v_{n}}
=\tilde{T}r \phi_{k}e^{H\cdot\phi+\Delta W}
\phi_{v_{1}}\phi_{v_{2}}\ldots \phi_{v_{n}}
\ee
and using the chain-rule for functional differentiation, we find
\be
\Sigma_{ij}
=\gamma_{ik}
\sum_{n=2}^{\infty}\frac{1}{n!}
{\cal G}_{kv_{1}v_{2}\ldots v_{n}}
\sigma_{u_{1}v_{1}}
\sigma_{u_{2}v_{2}}
\dots
\sigma_{u_{n}v_{n}}
G_{u_{1}u_{2}\ldots u_{n}\ell}\Gamma_{\ell j}
~~~.
\ee
Introduce the amputated vertices $\bar{\Gamma}$ and
$\bar{\gamma}$:
\be
{\cal G}_{kv_{1}v_{2}\ldots v_{n}}
=
{\cal G}_{km}
{\cal G}_{v_{1}w_{1}}
{\cal G}_{v_{2}w_{2}}
\ldots
{\cal G}_{v_{n}w_{n}}
\bar{\gamma}_{mw_{1}w_{2}\ldots w_{n}}
\ee
and
\be
G_{kv_{1}v_{2}\ldots v_{n}}
=
G_{km}
G_{v_{1}w_{1}}
G_{v_{2}w_{2}}
\ldots
G_{v_{n}w_{n}}
\bar{\Gamma}_{mw_{1}w_{2}\ldots w_{n}}
~~~.
\ee
Then we have quite generally
\be
\Sigma_{ij}
=\gamma_{ik}
\sum_{n=2}^{\infty}\frac{1}{n!}
{\cal G}_{km}
{\cal G}_{v_{1}w_{1}}
{\cal G}_{v_{2}w_{2}}
\ldots
{\cal G}_{v_{n}w_{n}}
\bar{\gamma}_{mw_{1}w_{2}\ldots w_{n}}
\nonumber
\ee
\be
\times
\sigma_{u_{1}v_{1}}
\sigma_{u_{2}v_{2}}
\dots
\sigma_{u_{n}v_{n}}
G_{u_{1}x_{1}}
G_{u_{2}x_{2}}
\ldots
G_{u_{n}x_{n}}
G_{p\ell}
\bar{\Gamma}_{p x_{1}x_{2}\ldots x_{n}}
\Gamma_{\ell j}
~~~.
\ee
We see factors of the FDRMP
\be
\bar{G}_{xw}=\sum_{uv}
{\cal G}_{vw}
\sigma_{uv}
G_{ux}
\ee
entering the development.  We can then write
\be
\Sigma_{ij}
=\sum_{n=2}^{\infty}\frac{1}{n!}
\bar{\gamma}_{i w_{1}w_{2}\ldots w_{n}}
\bar{G}_{x_{1}w_{1}}
\bar{G}_{x_{2}w_{2}}
\ldots
\bar{G}_{x_{n}w_{n}}
\bar{\Gamma}_{j x_{1}x_{2}\ldots x_{n}}
~~~.
\ee
In terms of Fourier transforms we have the general result:
\be
\Sigma_{\alpha_{i}\mu_{j}}(12)
=\sum_{n=2}^{\infty}\frac{1}{n!}\int d2d3 \ldots d(n+1)
\delta (1+2+\ldots+(n+1))
\nonumber
\ee
\be
\times
\bar{\gamma}^{*}_{\alpha_{i}\alpha_{1} \ldots \alpha_{n}}(12\ldots (n+1))
\bar{G}_{\alpha_{1}\mu_{1}}(1)
\bar{G}_{\alpha_{2}\mu_{2}}(2)
\ldots
\bar{G}_{\alpha_{(n+1)}\mu_{(n+1)}}(n+1)
\nonumber
\ee
\be
\times
\bar{\Gamma}_{\mu_{j}\mu_{1} \ldots \mu_{n}}(12\dots (n+1))
~~~.
\ee
Now we are interested in the $BB$ component
\be
\Sigma_{BB}(12)
=\sum_{n=2}^{\infty}\frac{1}{n!}\int d1 d2 \ldots dn
\delta (1+2+\ldots+(n+1))
\nonumber
\ee
\be
\times
\bar{\gamma}^{*}_{B\alpha_{1} \ldots \alpha_{(n+1)}}(12\ldots (n+1))
\bar{G}_{\alpha_{1}\mu_{1}}(1)
\bar{G}_{\alpha_{2}\mu_{2}}(2)
\ldots
\bar{G}_{\alpha_{(n+1)}\mu_{(n+1)}}(n+1)
\nonumber
\ee
\be
\times
\bar{\Gamma}_{B\mu_{1} \ldots \mu{(n+1)}}(12\ldots (n+1))
~~~.
\ee
In the long-time regime the internal propagators are dominated by
$\rho\rho$ lines and we have in the low-frequency regime
\be
\Sigma_{BB}(1)
=\sum_{n=2}^{\infty}\frac{1}{n!}\int d2 d3 \ldots d(n+1)
\delta (1+2+\ldots+(n+1))
\nonumber
\ee
\be
\times
\bar{\gamma}^{*}_{B\rho \ldots \rho}(23\ldots (n+1))
\bar{G}_{\rho\rho}(2)
\bar{G}_{\rho\rho}(3)
\ldots
\bar{G}_{\rho\rho}(n+1)
\bar{\Gamma}_{B\rho \ldots \rho}(23\ldots (n+1))
~~~.
\ee
Next we note that the major contribution to the integals is from the
low-frequency portions of the $\bar{G}_{\rho\rho}$. In that regime we can
evaluate the vertices at zero-frequency.  At zero-frequency we can use
the vertex theorems (see Appendix A)
\be
\bar{\gamma}_{B\rho \ldots \rho}(q_{1},0,k_{2},0\ldots k_{n+1},0)
=-\frac{1}{\beta \bar{\rho}^{n}}
\ee
\be
\bar{\Gamma}_{B\rho \ldots \rho}(q_{1},0,k_{2},0\ldots k_{n+1},0)
=-\beta^{-1}\bar{\gamma} (q_{1},k_{2},\ldots ,k_{n+1})
\ee
to obtain
\be
\beta^{2}\Sigma_{BB}(1)
=\sum_{n=2}^{\infty}\frac{1}{n!}\int d2 d3 \ldots d(n+1)
\delta (1+2+\ldots+(n+1))
\nonumber
\ee
\be
\times
\frac{1}{\bar{\rho}^{n}}
\bar{G}_{\rho\rho}(2)
\bar{G}_{\rho\rho}(3)
\ldots
\bar{G}_{\rho\rho}(n+1)
\bar{\gamma}(q_{1},k_{2}\ldots k_{n+1})
~~~.
\ee
This is the long-time approximation for the collective
part of the memory function. If we ignore vertex corrections
and replace the static vertices with its ideal-gas form
\be
\gamma (q_{1}k_{2}\ldots k_{n+1})
=\frac{1}{\bar{\rho}^{n}}
\ee
we obtain
\be
\beta^{2}\Sigma_{BB}(1)
=\sum_{n=2}^{\infty}\frac{1}{n!}\int d2 d3 \ldots d(n+1)
\delta (1+2+\ldots+(n+1))
\nonumber
\ee
\be
\times
\frac{1}{\bar{\rho}^{2n}}
\bar{G}_{\rho\rho}(2)
\bar{G}_{\rho\rho}(3)
\ldots
\bar{G}_{\rho\rho}(n+1)
~~~.
\ee
If we express the $\bar{G}_{\rho\rho}(k_{i},\omega_{i})$
in terms of its space-time Fourier transform we can do the
frequency integrals and obtain the very simple result
\be
\beta^{2}\Sigma_{BB}(1)
=\sum_{n=2}^{\infty}\frac{1}{n!}\int dt_{1}e^{it_{1}\omega_{1}}
\int d^{d}r_{1}e^{ir_{1}\cdot q_{1}}(-D(r_{1},t_{1}))^{n}
\ee
where we have
\be
D(r_{1},t_{1})=-\frac{\bar{G}_{\rho\rho}(r_{1},t_{1})}
{\bar{\rho}^{2}}
~~~.
\ee
We can do the "loop" sum to obtain
\be
\beta^{2}\Sigma_{BB}(1)
=\int dt_{1}e^{it_{1}\omega_{1}}
\int d^{d}r_{1}e^{ir_{1}\cdot q_{1}}H[D(r_{1},t_{1})]
\ee
where
\be
H[x]=e^{-x}-1+x
~~~.
\ee
If we expand in powers of $D$ and keep the lowest nonzero result
we obtain our previous perturbative results at second order\cite{VC}.
We are able to sum up all the loop diagrams, wihout vertex corrections,
and go well beyond
MCT.

We also assume, consistent with the long-time approximation, that
for low-frequencies
\be
G_{\rho\rho}(q,\omega )\gg G_{\rho B}(q,\omega)
\ee
and in $\bar{G}_{\rho\rho}(q, \omega )$ the
$G_{\rho\rho}(q, \omega )$ term dominates and its coefficient
can be replaced by its $\omega =0$ value.  Then using the
FDR we find
\be
\frac{\bar{G}_{\rho\rho}(q, \omega )}{\bar{\rho}^{2}}
=-\beta\bar{\rho}V(q)G_{\rho\rho}(q, \omega)/\bar{\rho}^{2}
\nonumber
\ee
\be
=-\bar{V}(q)\bar{S}(q)F(q, \omega)/\bar{\rho}
=-D(q,\omega )
\ee
where
\be
\bar{V}(q)=\beta\bar{\rho}V(q)
~~~,
\ee
\be
G_{\rho\rho}(q, \omega)=S(q)F(q,\omega )
~~~,
\ee
and
\be
S(q)=\bar{\rho}\bar{S}(q)
~~~.
\ee

Taking the inverse Fourier transforms we obtain the very simple
result
\be
\beta^{2}\Sigma_{BB}(r,t)
=H[D(r,t)]
~~~.
\label{eq:34a}
\ee

Finally, in conventional MCT the ENE problem is formulated entirely
in terms of the structure factor.  In our formulation here
this is true only if we ignore vertex corrections.  In general
one must also include the $n$-point vertex corrections at $n$-loop
order.  This will be persued elsewhere.

\section{Analytic Long-time analysis}

Putting the Fourier-Laplace transform of Eq.(\ref{eq:34a}) into
Eq.(\ref{eq:171}),
we find that the universal long-time kinetics is governed in the
no-vertex  correction approximation by the equation:
\be
\frac{F(q,z)}{1-zF(q,z)}
=S(q)\int d^{3}r e^{iq\cdot r}\int_{0}^{\infty}dt e^{-zt}H[D(r,t)]
\label{eq:ana}
\ee
where
\be
H[x]=e^{-x}-1+x
\ee
and
\be
D(r,t)=\frac{1}{\bar{\rho}}\int\frac{d^{3}q}{(2\pi )^{3}}
e^{-iq\cdot r}\bar{V}(q)\bar{S}(q)F(q,t)
~~~.
\ee
Fourier-transforming over space
\be
D(q,t)=\frac{1}{\bar{\rho}}\bar{V}(q)\bar{S}(q)F(q,t)
\ee
or Laplace transforming over time
\be
D(r,z)=\frac{1}{\bar{\rho}}\int\frac{d^{3}q}{(2\pi )^{3}}
e^{-iq\cdot r}\bar{V}(q)\bar{S}(q)F(q,z)
\ee
or the double transform
\be
D(q,z)=\frac{1}{\bar{\rho}}\bar{V}(q)\bar{S}(q)F(q,z)
~~~.
\ee

If we replace $H[x]$ with $\frac{1}{2}x^{2}$ we have the ergodic-nonergodic
problem studied in SDENE, ND2 and SM.  Using the full $H[x]$ we have the
ENEP where we keep all loops.  Note the ENEP is the same for SD and ND.

Now we look for a solution of Eq.(\ref{eq:ana}) of the form
\be
F(r,t)=f(r)+\psi (r,t)
\ee
or
\be
F(q,t)=f(q)+\psi (q,t)
\ee
valid at long times and where $\psi (r,t)$ is "small".
$f(q)$ is the crucial nonergodicity parameter.

In terms of
the Laplace-transform over time
\be
F(r,z)=\frac{f(r)}{z}+\psi (r,z)
\ee
and the Fourier transform over space
\be
F(q,z)=\frac{f(q)}{z}+\psi (q,z)
~~~.
\ee
The quantity $D(r,t)$ can be written
\be
D(r,t)=\frac{1}{\bar{\rho}}
\int\frac{d^{3}q}{(2\pi )^{3}}e^{-iq\cdot r}\bar{V}(q)\bar{S}(q)
\left[f(q)+\psi (q,t)\right]
\nonumber
\ee
\be
=D_{0}(r)+\Delta (r,t)
\ee
where
\be
D_{0}(r)=\frac{1}{\bar{\rho}}\int\frac{d^{3}q}{(2\pi )^{3}}
e^{-iq\cdot r}\bar{V}(q)\bar{S}(q)
f(q)
\ee
and
\be
\Delta (r,t)=\frac{1}{\bar{\rho}}\int\frac{d^{3}q}{(2\pi )^{3}}
e^{-iq\cdot r}\bar{V}(q)\bar{S}(q)
\psi (q,t)
\label{eq:148}
\ee
or
\be
\Delta (r,z)=\frac{1}{\bar{\rho}}\int\frac{d^{3}q}{(2\pi )^{3}}
e^{-iq\cdot r}\bar{V}(q)\bar{S}(q)
\psi (q,z)
~~~.
\label{eq:149}
\ee
Expanding in powers of $\Delta$ we find keeping terms of second order
\be
H[D(r,t)]=H[D_{0}(r)+\Delta (r,t)]
\nonumber
\ee
\be
=H[D_{0}(r)]+\Delta (r,t)(1-e^{-D_{0}(r)})
+\frac{1}{2}\Delta ^{2}(r,t)e^{-D_{0}(r)}
~~~.
\ee
This is inserted into the right-hand-side of Eq.(\ref{eq:ana}).
The left-hand side of the same equation takes the expanded form
\be
L(q,z)=\frac{F(q,z)}{1-zF(q,z)}
\nonumber
\ee
\be
=\frac{f(q)}{z(1-f(q))}
+\frac{\psi (q,z)}{(1-f(q))^{2}}
+\frac{z\psi^{2} (q,z)}{(1-f(q))^{3}}
~~~.
\ee
Eq.(\ref{eq:ana}) takes the form
\be
L(q,z)
=R_{1}(q,z)
+R_{2}(q,z)
+R_{3}(q,z)
\ee
where
\be
R_{1}(q,z)
=S(q)\frac{H_{0}(q)}{z}
\ee
with
\be
H_{0}(q)=\int d^{3}r e^{iq\cdot r}
H[D_{0}(r)]
\ee
and
\be
R_{2}(q,z)
=S(q)\int d^{3}r e^{iq\cdot r}
\Delta (r,z)(1-e^{-D_{0}(r)})
\label{eq:155}
\ee
and
\be
R_{3}(q,z)
=S(q)\int d^{3}r e^{iq\cdot r}\int_{0}^{\infty}dt e^{-zt}
\frac{1}{2}\Delta ^{2}(r,t)e^{-D_{0}(r)}
~~~.
\label{eq:156}
\ee
Now we match orders in our perturbation theory. The leading order goes as
$1/z$ for small $z$.  Matching coefficients at leading order
we have
\be
\frac{f(q)}{1-f(q)}=S(q)H_{0}(q)
\label{eq:157}
\ee
and the higher-order contributions are given by
\be
\frac{\psi (q,z)}{(1-f(q))^{2}}
+\frac{z\psi^{2} (q,z)}{(1-f(q))^{3}}
=R_{2}(q,z)+R_{3}(q,z)
~~~.
\label{eq:HO}
\ee
Eq.(\ref{eq:157}) is critical in the development.  The question whether
we have an nonergodic phase is determined\cite{RM} by this "static" quantity.
A nonzero $f$ depends on static parameters.

Let us focus on the higher-order contributions to Eq.(\ref{eq:HO}).
Consider the linear
contribution. Using Eq.(\ref{eq:155}) and Eq.(\ref{eq:149}), we have
\be
R_{2}(q,z)
=\int\frac{d^{3}k}{(2\pi )^{3}}
\bar{C}_{q,k}\psi (k,z)
\ee
where we have defined the important matrix,
\be
\bar{C}_{q,k}
=S(q)\frac{1}{\bar{\rho}}\int d^{3}r e^{iq\cdot r}
(1-e^{-D_{0}(r)})
e^{-ik\cdot r}\bar{V}(k)\bar{S}(k)
\nonumber
\ee
\be
=\bar{S}(q)M(q-k)\bar{V}(k)\bar{S}(k)
\ee
and where we have
\be
M(q)
=\int d^{3}r e^{iq\cdot r}
(1-e^{-D_{0}(r)}).
\ee
The higher-order contributions, Eq.(\ref{eq:HO}), can now be written
\be
\frac{\psi (q,z)}{(1-f(q))^{2}}
+\frac{z\psi^{2} (q,z)}{(1-f(q))^{3}}
=\int\frac{d^{3}k}{(2\pi )^{3}}
\bar{C}_{q,k}\psi (k,z)
+R_{3}(q,z)
~~~.
\label{eq:HO1}
\ee
It is notationally advantageous to introduce
\be
\psi (q,z)=(1-f(q))^{2}\phi (q,z)
\ee
and find in terms of $\phi$
\be
\phi (q,z)
+z(1-f(q))\phi^{2} (q,z)
=\int\frac{d^{3}k}{(2\pi )^{3}}
C_{q,k}\phi (k,z)
+R_{3}(q,z)
\label{eq:163}
\ee
where
\be
C_{q,k}
=\bar{S}(q)M(q-k)\bar{V}(k)\bar{S}(k)
(1-f(k))^{2}
~~~.
\label{eq:164}
\ee

Let us turn to $R_{3}(q,z)$. Starting with Eqs.(\ref{eq:156}) and (\ref{eq:148})
we have:
\be
R_{3}(q,z)
=S(q)\int d^{3}r e^{iq\cdot r}\int_{0}^{\infty}dt e^{-zt}
\frac{1}{2}\Delta ^{2}(r,t)e^{-D_{0}(r)}
\ee
and
\be
\Delta (r,t)=\frac{1}{\bar{\rho}}\int\frac{d^{3}k}{(2\pi )^{3}}
e^{-ik\cdot r}\bar{V}(k)\bar{S}(k)
(1-f(k))^{2}\phi (k,t)
\ee
we can write
\be
R_{3}(q,z)=
\int\frac{d^{3}k}{(2\pi )^{3}}
\int\frac{d^{3}p}{(2\pi )^{3}}
C_{q,k,p}
\int_{0}^{\infty}dt e^{-zt}\phi (k,t)\phi (p,t)
\ee
where
\be
C_{q,k,p}
=\frac{1}{\bar{\rho}^{2}}\frac{S(q)}{2}\int d^{3}r e^{i(q-k-p)\cdot r}
e^{-D_{0}(r)}
\bar{V}(k)\bar{S}(k)(1-f(k))^{2}
\bar{V}(p)\bar{S}(p)(1-f(p))^{2}
~~~.
\label{eq:168}
\ee
Pulling all of this together we can rewrite Eq.(\ref{eq:HO1}) as
\be
\int\frac{d^{3}k}{(2\pi )^{3}}
(\delta_{q,k}-C_{q,k})\phi (k,z)
+z(1-f(q))\phi^{2} (q,z)
\nonumber
\ee
\be
=\int\frac{d^{3}k}{(2\pi )^{3}}
\int\frac{d^{3}p}{(2\pi )^{3}}
C_{q,k,p}
\int_{0}^{\infty}dt e^{-zt}\phi (k,t)\phi (p,t)
~~~.
\ee
Continuing our perturbation theory and matching equal powers, we assume
\be
\phi (q,t)=\phi^{(1)}(q,t)+
\phi^{(2)}(q,t)
\ee
and require
that the first-order terms must satisfy
\be
\int\frac{d^{3}k}{(2\pi )^{3}}
(\delta_{q,k}-C_{q,k})\phi^{(1)} (k,z)=0
~~~.
\ee
Then
at second order we have
\be
\int\frac{d^{3}k}{(2\pi )^{3}}
(\delta_{q,k}-C_{q,k})\phi ^{(2)} (k,z)
+z(1-f(q))[\phi^{(1)}(q,z)]^{2}
\nonumber
\ee
\be
=\int\frac{d^{3}k}{(2\pi )^{3}}
\int\frac{d^{3}p}{(2\pi )^{3}}
C_{q,k,p}
\int_{0}^{\infty}dt e^{-zt}\phi^{(1)} (k,t)\phi ^{(1)}(p,t)
~~~.
\ee
Looking for a first-order solution of the form
\be
\phi^{(1)}(q,t)=A\epsilon_{q}\phi_{\nu}(t)
\ee
where $A$ is independent of $q$ and $t$, and $\epsilon_{p}$
is the right eigenfunction of $C_{q,k}$  with unit eigenvalue
\be
\int\frac{d^{3}k}{(2\pi )^{3}}
(\delta_{q,k}-C_{q,k})\epsilon_{k}=0
~~~.
\label{eq:174}
\ee
We also introduce the left eigenfunction
\be
\int\frac{d^{3}k}{(2\pi )^{3}}
(\delta_{q,k}-C_{k,q})\hat{\epsilon}_{k}=0
~~~.
\label{eq:175}
\ee
We normalize the eigenfunctions using
\be
\int\frac{d^{3}k}{(2\pi )^{3}}
\hat{\epsilon}_{k}
\epsilon_{k}=1
\ee
and
\be
\int\frac{d^{3}k}{(2\pi )^{3}}
\hat{\epsilon}_{k}(1-f(k))
\epsilon^{2}_{k}=1
~~~.
\ee
The second-order contribution is then given by
\be
\int\frac{d^{3}k}{(2\pi )^{3}}
(\delta_{q,k}-C_{q,k})\phi ^{(2)} (k,z)
+z(1-f(q))[A\epsilon_{q}\phi_{\nu}(z)]^{2}
\nonumber
\ee
\be
=\int\frac{d^{3}k}{(2\pi )^{3}}
\int\frac{d^{3}p}{(2\pi )^{3}}
C_{q,k,p}\epsilon_{k}\epsilon_{p}A^{2}
\int_{0}^{\infty}dt e^{-zt}\phi_{\nu}^{2} (t)
~~~.
\ee
Matrix multiply by $\hat{\epsilon}_{q}$. This kills the first term on the
left and after using the normalizations gives the rather simple result:
\be
zA^{2}\phi_{\nu}^{2}(z)=A^{2}\lambda
\int_{0}^{\infty}dt e^{-zt}\phi_{\nu}^{2} (t)
\label{eq:180}
\ee
where we introduce the parameter
\be
\lambda=
\int\frac{d^{3}q}{(2\pi )^{3}}
\int\frac{d^{3}k}{(2\pi )^{3}}
\int\frac{d^{3}p}{(2\pi )^{3}}
\hat{\epsilon}_{q}
C_{q,k,p}\epsilon_{k}\epsilon_{p}
\ee
and Eq.(\ref{eq:180}) reduces to
\be
z\phi_{\nu}^{2}(z)=\lambda
\int_{0}^{\infty}dt e^{-zt}\phi_{\nu}^{2} (t)
\label{eq:651}
~~~.
\ee

Eq.(\ref{eq:651}) is satisfied by a
power-law solution
\be
\phi_{\nu} (t) =t^{-a}
\ee
with Laplace transform
\be
\phi_{\nu} (z) =\frac{\Gamma (1-a)}{z^{1-a}}
~~~.
\ee
Putting this into Eq.(\ref{eq:651}) we find
\be
z\left(\frac{\Gamma (1-a)}{z^{1-a}}\right)^{2}=
\lambda \frac{\Gamma (1-2a)}{z^{1-2a}}
\ee
or, the MCT result,
\be
\frac{\Gamma^{2}(1-a)}{\Gamma (1-2a)}=\lambda
~~~.
\label{eq:185}
\ee

As discussed in SM, one can go further and treat the full
two-step decay process including the von Schweidler\cite{vS} contribution.

\section{Pseudo-potential}

In SDENE we worked to second order in the pseudo-potential
$V(q)$ which was expressed in terms of the structure factor $S(q)$.
The development was self-consistent and thorough.  It involved
the construction of the second-order self energies
$\Sigma_{B\rho}^{(L,2)}$ and
$\Sigma_{\rho B}^{(L,2)}$ , showing that these satisfy the appropriate FDR
and the associated static sum-rule giving the cooresponding static
approximation in SDENE.

In the present case it is easier to simply work out the static manipulations
giving the direct-correlation functions, $C_{D}(q)$,
 as a power-series in the static
pseudo-potential.  In the all-loops approximation we ignore vertex
corrections and do the loop sums to obtain the static approximation
\be
-\rho C_{D}(q)=\bar{V}(q)
-\frac{24\eta}{q}\int_{0}^{\infty }rdr \sin(qr)H(Y(r))
\ee
\be
Y(r)=\frac{1}{12\pi\eta r}
\int_{0}^{\infty}qdq \sin (qr)\bar{V}(q)\bar{S}(q)
\ee
and
\be
H(x)=e^{-x}-1+x
~~~.
\ee

\section{ENE Problem}

One then has the ENE problem to solve.  Given a static structure factor,
is there a solution to Eq.(\ref{eq:157}) with $f(q)\neq 0$.  This requires
a numerical treatment of Eq.(\ref{eq:157}).  For those cases where we
find a nonergodic regime, $f(q)\neq 0$, we must solve the eigenvalue
problems posed by Eq.(\ref{eq:174}) and Eq.(\ref{eq:175})
with the matrix $C_{q,k}$ defined by Eq.(\ref{eq:164}).
Then one is in a position to compute the parameter $\lambda$ given by
Eqs.(\ref{eq:180}) and (\ref{eq:168}).
The power-law time-decay exponent is given the by Eq.(\ref{eq:185}).
This numerical analysis has been carried out in the simplest case of
hard-spheres and using the Percus-Yevic approximation for the
static structure factor. An ENE transition is found to occur for
reduced density $\eta^{*}=0.651...$.  A full numerical
analysis will be given in
a separate paper.

\section{Summary}

We have shown here how one can the study of the ENE transition
for dense fluids well beyond the one-loop treatment of mode coupling
theory.  We have shown how one can treat multiple-loop
calculations in a straight-forward manner.
We focussed on the approximation where we ignore static vertex
corrections and sum all loop contributions.
Still left to be worked out is the role of vertex corrections.
Do they change the picture developed here in any significant way?
This can be tested by doing perturbation theory in the pseudo-potential.
This will be discussed elsewhere.

A key result is that we have shown that SD and ND share the same
ENE problem.  It would be interesting to investigate whether the
more elaborate Fokker-Planck dynamics falls into the same universality
class.  We speculate that it would lead  to the same ENE problem.
As a first step one would need to show that the core-variables
$\rho$ and $B$ satisfy the same FDRs as for SD and ND.

The author thanks Professor S. Das,
David McCowan and Paul Spyridis for
comments and help with the manuscript.

\newpage

\centerline{Appendices}

\section{Higher-Order Thermodynamic Sum Rules}

\subsection{B-Theorem}

In this section we show how  the fluctuation-dissipation relations (FDR)
can be used to derive thermodynamic sum rules.
We start with the general FDR
for the core fields
$\rho (x,t)$ and  $B(x,t)$.
The n-point cumulants in terms of time-Fourier
transforms obey
\be
G_{BB\ldots B\rho\ldots\rho}(12\ldots\ell,\ell+1\ldots n)
=G^{*}_{\psi\psi...\psi\rho\ldots\rho}(12\ldots\ell,\ell+1\ldots n)
\label{eq:190}
\ee
where
\be
\psi_{i}=B_{i}-i\beta\omega_{i}\rho_{i}
~~~.
\ee

Let us now consider the case of $n$ $B$-variables and one $\rho$ variable.
We can write
\be
G_{\rho B\ldots B}(12\ldots n+1)
=G^{*}_{\rho\psi\ldots\psi} (12\ldots n+1)
\label{eq:21}
~~~.
\ee
Let us introduce a little notation.
If we remember the conserving frequency $\delta$-function we can write
\be
G_{\rho B\ldots B}(12\ldots n+1)=
\tilde{G}_{\rho B\ldots B}
(\omega_{1},\omega_{2},\ldots ,\omega_{n},\omega_{n+1})
\delta (\omega_{1}+\omega_{2}+\ldots +\omega_{n+1})
\label{eq:543}
\ee
where we surpress the wavenumber dependence.
Using this notation in Eq.(\ref{eq:190})
and setting $\omega_{2}=\omega_{3}=\ldots =\omega_{n}=0$,
we have $\omega_{n+1}=-\omega_{1}$,
and
\be
\tilde{G}_{\rho B\ldots B}(\omega_{1},0,\ldots ,0,-\omega_{1})
=\tilde{G}^{*}_{\rho B\ldots B}(\omega_{1},0,\ldots ,0,-\omega_{1})
-i\beta \omega_{1}
\tilde{G}^{*}_{\rho B\ldots\rho}(\omega_{1},0,\ldots ,0,-\omega_{1})
\ee
or
\be
2i\tilde{G}^{''}_{\rho B\ldots B}(\omega_{1},0,\ldots ,0,-\omega_{1})
=-i\beta \omega_{1}
\tilde{G}^{*}_{\rho B\ldots\rho}(\omega_{1},0,\ldots ,0,-\omega_{1})
~~~.
\ee
Notice that
\be
\tilde{G}^{''}_{\rho B\ldots B\rho}(\omega_{1},0,\ldots ,0,-\omega_{1}) =0
~~~.
\ee
Now consider the Fourier representation:
\be
G_{\rho B\ldots B}(t_{1},t_{2},\ldots ,t_{n+1})
=\int\frac{d\omega_{1}}{2\pi}
e^{-i\omega_{1}(t_{1}-t_{n+1})}
\int\frac{d\omega_{2}}{2\pi}
e^{-i\omega_{2}(t_{2}-t_{n+1})}
\cdots
\nonumber
\ee
\be
\times
\int\frac{d\omega_{n}}{2\pi}
e^{-i\omega_{n}(t_{n}-t_{n+1})}
\tilde{G}_{\rho B\ldots B}(\omega_{1},\omega_{2},\ldots,\omega_{n+1})
\ee
where
$\omega_{n+1}=-\omega_{1}-\omega_{2}-\ldots -\omega_{n}$.  Now if
$t_{n+1}> t_{1}$ then
$G_{\rho B\ldots B}(t_{1},t_{2},\ldots ,t_{n+1})=0$.  This implies that
$\tilde{G}_{\rho B\ldots B}(\omega_{1},\omega_{2},\ldots,\omega_{n+1})$
is analytic in the upper-half plane.  So
$\tilde{G}_{\rho B\ldots B B}(\omega_{1},0,\ldots,0,-\omega_{1})$
is analytic in the upper half plane.
We can then construct the dispersion relation
\be
\tilde{G}_{\rho B\ldots B B}(\omega_{1},0,\ldots ,0,-\omega_{1})
=\int\frac{d\bar{\omega}}{\pi}\frac{\tilde{G}^{''}_{\rho B\ldots B B}
(\bar{\omega},0,\ldots,0,-\bar{\omega})}{\bar{\omega}-\omega_{1}-i\eta}
~~~.
\ee
We showed earlier
\be
\tilde{G}^{''}_{\rho B\ldots B}(\omega_{1},0,\ldots ,0,-\omega_{1})
=-\frac{\beta \omega_{1}}{2}
\tilde{G}^{*}_{\rho B\ldots\rho}(\omega_{1},0,\ldots ,0,-\omega_{1})
\ee
so we have
\be
\tilde{G}_{\rho B\ldots B B}(\omega_{1},0,\ldots,0,-\omega_{1})
=-\int\frac{d\bar{\omega}}{2\pi}
\frac{\beta\bar{\omega}\tilde{G}^{*}_{\rho B\ldots B \rho}
(\bar{\omega},0,\ldots ,0,-\bar{\omega})}{\bar{\omega}-\omega_{1}-i\eta}
~~~.
\ee

Taking $\omega_{1}=0$ we find
\be
\tilde{G}_{\rho B\ldots B B}(0,0,\ldots ,0,0)
=-\int\frac{d\bar{\omega}}{2\pi}\beta\tilde{G}^{*}_{\rho B\ldots B \rho}
(\bar{\omega},0,\ldots ,0,-\bar{\omega})
~~~.
\label{eq:A1}
\ee

Returning to the time domain we have
\be
G_{\rho B\ldots \rho}(t_{1},t_{2},\ldots ,t_{n+1})
=\int\frac{d\omega_{1}}{2\pi}
e^{-i\omega_{1}(t_{1}-t_{n+1})}
\int\frac{d\omega_{2}}{2\pi}
e^{-i\omega_{2}(t_{2}-t_{n+1})}
\cdots
\nonumber
\ee
\be
\times
\int\frac{d\omega_{n}}{2\pi}
e^{-i\omega_{n}(t_{n}-t_{n+1})}
\tilde{G}_{\rho B\ldots \rho}(\omega_{1},\omega_{2},\ldots,\omega_{n+1})
~~~.
\ee
Fourier transforming over $t_{2},\ldots,t_{n}$ we find
\be
G_{\rho B\ldots B\rho}(t_{1},\omega_{2},\ldots \omega_{n},t_{n+1})=
\nonumber
\ee
\be
\int\frac{d\omega_{1}}{2\pi}
e^{-i\omega_{1}(t_{1}-t_{n+1})}
e^{it_{n+1}(\omega_{2}+\ldots +\omega_{n})}
\tilde{G}_{\rho B\ldots B\rho}(\omega_{1},0,\ldots,0,-\omega_{1})
~~~.
\ee
Taking all the external frequencies to zero we obtain
\be
G_{\rho B\ldots B\rho}(t_{1},0,\ldots ,0,t_{n+1})
=\int\frac{d\omega_{1}}{2\pi}
e^{-i\omega_{1}(t_{1}-t_{n+1})}
\tilde{G}_{\rho B\ldots B\rho}(\omega_{1},0,\ldots,0,-\omega_{1})
~~~.
\ee
Setting $t_{n+1}=t_{1}$ we obtain the important result
\be
G_{\rho B\ldots B\rho}(t_{1},0,\ldots ,0,t_{1})
=\int\frac{d\omega_{1}}{2\pi}
\tilde{G}_{\rho B\ldots B\rho}(\omega_{1},0,\ldots,0,-\omega_{1})
~~~.
\ee
Using this result in Eq.(\ref{eq:A1}) we find
\be
\tilde{G}_{\rho B\ldots B B}(0,0,\ldots,0,0)
=-\beta
G_{\rho B\ldots B\rho}(t_{1},0,\ldots ,0,t_{1})
~~~.
\ee
If we rewrite this in the form
\be
\tilde{G}_{\rho B\ldots B B}(0,0,\ldots,0,0)
=-\beta
G_{\rho\rho B\ldots B}(t_{1},t_{1},0,\ldots ,0)
~~~.
\ee
It is easy to see that one can turn the crank and find
\be
G_{\rho\rho B\ldots B}(t_{1},t_{1},0,\ldots ,0)
=-\beta
G_{\rho\rho \rho B\ldots B}(t_{1},t_{1},t_{1},0,\ldots ,0)
~~~.
\ee
and
\be
\tilde{G}_{\rho B\ldots B B}(0,0,\ldots,0,0)
=(-\beta )^{n}
G_{\rho \rho \ldots \rho}(t_{1},t_{1},\ldots ,t_{1})
\nonumber
\ee
\be
=(-\beta )^{n} S_{C}(q_{1},q_{2},\ldots ,q_{n+1})
\label{eq:A20}
\ee
where $S_{C}$ is the $n+1$ static density cumulant.

\subsection{n-point Amputated Vertices}

We define $n$-point amputated vertices with
\be
G_{\alpha_{1}\alpha_{2}\ldots \alpha_{n}}(12\ldots n)
=G_{\alpha_{1}\mu_{1}}(1)
G_{\alpha_{2}\mu_{2}}(2)
\cdots
G_{\alpha_{n}\mu_{n}}(n)
\bar{\Gamma}_{\mu_{1}\mu_{2}\ldots\mu_{n}}(12\ldots n)
~~~.
\ee
Notice that
\be
G_{BB\ldots B}(12\ldots n)
=G_{B\rho }(1)
G_{B\rho }(2)
\cdots
G_{B\rho }(n)
\bar{\Gamma}_{\rho_{1}\rho_{2}\ldots\rho_{n}}(12\ldots n)=0
~~~.
\ee
We are interested in
\be
G_{\rho B\ldots B}(12\ldots n)
=G_{\rho B}(1)
G_{B\rho}(2)
\cdots
G_{B\rho}(n)
\bar{\Gamma}_{B\rho\ldots\rho}(12\ldots n)
~~~.
\ee
Setting all external frequencies to zero and using Eq.(\ref{eq:A20})
cancelling
the common factors of $-\beta$ we find the very useful result
\be
-\beta \bar{\Gamma}_{B\rho\ldots\rho}(0,0,\ldots ,0;q_{1},q_{2},\ldots ,q_{n})
=\bar{\gamma}_{\rho\rho\ldots\rho}(q_{1},q_{2},\ldots ,q_{n})
\ee
where $\bar{\gamma}$ is the static $n$-point amputated vertex.

\subsection{Single-particle Quantities}

It should be noted that this analysis goes through for
the single-particle quantities.  Thus we can replace
$G_{\rho BB}\rightarrow {\cal G}_{\rho BB}$.
Then we have the identities
\be
G_{\rho }(x_{1},t_{1})
=Tr \phi_{\rho}(x_{1},t_{1})
e^{\Delta W}=\bar{\rho}
\ee
\be
{\cal G}_{\rho B}(x_{1},x_{2};0)
=-\beta Tr \phi_{\rho}(x_{1},t_{1})
 \phi_{\rho}(x_{2},t_{1})
e^{\Delta W}
\nonumber
\ee
\be
=-\beta \bar{\rho}\delta (x_{1}-x_{2})
~~~.
\ee
\be
\tilde{{\cal G}}_{\rho BB}(x_{1},x_{2},x_{3};0,0,0)
=\beta^{2} Tr \phi_{\rho}(x_{1},t_{1})
 \phi_{\rho}(x_{2},t_{1})
 \phi_{\rho}(x_{3},t_{1})
e^{\Delta W}
\nonumber
\ee
\be
=\beta^{2} \bar{\rho}\delta (x_{1}-x_{2})
\delta (x_{1}-x_{3})
\ee
\be
\tilde{{\cal G}}_{\rho BBB}(x_{1},x_{2},x_{3},x_{4};0,0,0,0)
=-\beta^{3} Tr \phi_{\rho}(x_{1},t_{1})
 \phi_{\rho}(x_{2},t_{1})
 \phi_{\rho}(x_{3},t_{1})
 \phi_{\rho}(x_{4},t_{1})
e^{\Delta W}
\nonumber
\ee
\be
=-\beta^{3} \bar{\rho}\delta (x_{1}-x_{2})
\delta (x_{1}-x_{3})
\delta (x_{1}-x_{4})
\ee
an so on.  Fourier transforming over space:
\be
{\cal G}_{\rho B}(q_{1},q_{2};0)
=-\beta \bar{\rho}\delta (q_{1}+q_{2})
\ee
\be
\tilde{{\cal G}}_{\rho BB}(q_{1},q_{2},q_{3};0,0,0)
=\beta^{2} \bar{\rho}\delta (q_{1}+q_{2}+q_{3})
\ee
\be
\tilde{{\cal G}}_{\rho BBB}(q_{1},q_{2},q_{3},q_{4};0,0,0,0)
=-\beta^{3} \bar{\rho}\delta (q_{1}+q_{2}+q_{3}+q_{4})
\ee
and so on.
Then, for the associated amputated vertices,
\be
\gamma_{B\rho}(q_{1};0)=-\frac{1}{\beta\bar{\rho}}
\ee
\be
\gamma_{B\rho\rho}(q_{1},q_{2},q_{3};0,0,0)
=\frac{1}{\beta\bar{\rho}^{2}}
\ee
\be
\bar{\gamma}_{B\rho\rho\rho}(q_{1},q_{2},q_{3},q_{4};0,0,0,0)
=-\frac{1}{\beta\bar{\rho}^{3}}
~~~.
\ee

\section{FDR Matrix Propagators}

A FDR matrix propagator (FDRMP) $A_{\mu\nu}(q,\omega )$
satisfies the properties
\be
A_{\mu\nu}(q,\omega )
=A_{\nu\mu}^{*}(q,\omega )
\ee
\be
+i\beta \omega A_{\rho\rho}(q,\omega )
=A_{B\rho}(q,\omega )-A_{\rho B}(q,\omega )
\ee
\be
A_{\rho B}(q,\omega )=\int\frac{d\bar{\omega}}{2\pi}
\frac{\beta\bar{\omega}A_{\rho\rho}(q,\bar{\omega})}
{\omega -\bar{\omega}+i\eta}
\ee
and
\be
A_{\rho B}(q,0)=\int\frac{d\bar{\omega}}{2\pi}
\beta A_{\rho\rho}(q,\bar{\omega})
\ee
with
\be
A_{BB}(q,\omega )=0
~~~.
\ee
If $A_{\alpha\beta}(q,\omega )$ and
$C_{\alpha\beta}(q,\omega )$ are  FDR matrix
propagators  then
\be
D_{\alpha\beta}(q,\omega)=
\sum_{\mu\nu}A_{\alpha\mu}(q,\omega)\sigma_{\mu\nu}(q)
C_{\nu\beta}(q,\omega)
\ee
is also a FDR matrix propagator.
The proof is rather direct.  Look first at the response channel:
\be
D_{BB}(q,\omega)=
\sum_{\mu\nu}A_{B\mu}(q,\omega)\sigma_{B\nu}(q)
C_{\nu\beta}(q,\omega)
\nonumber
\ee
\be
=A_{B\rho}(q,\omega)\sigma_{\rho\rho}(q)
C_{\rho B}(q,\omega)=0
~~~.
\ee
Consider
next the off -diagonal componets
\be
D_{\rho B}(q,\omega)=
A_{\rho B}(q,\omega)\sigma_{B\rho}(q)
C_{\rho B}(q,\omega)
\nonumber
\ee
\be
=A_{\rho B}(q,\omega)V(q)
C_{\rho B}(q,\omega)
\ee
\be
D_{B\rho }(q,\omega)=
A_{B\rho }(q,\omega)V(q)
C_{B\rho}(q,\omega)
\ee
It is easy to see that
\be
D_{\rho B}(q,\omega)=
D^{*}_{B\rho }(q,\omega)
~~~.
\ee
Next consider the diagonal component
\be
D_{\rho \rho}(q,\omega)=
A_{\rho \rho}(q,\omega)V(q)C_{B\rho}(q,\omega )
+A_{\rho B}(q,\omega)V(q)C_{\rho \rho}(q,\omega )
\nonumber
\ee
\be
=\frac{V(q)}{i\beta\omega}\left[(A_{B\rho}(q,\omega)
-A_{\rho B}(q,\omega))C_{B\rho}(q,\omega )
+A_{\rho B}(q,\omega)
(C_{B\rho}(q,\omega )
-C_{\rho B}(q,\omega ))\right]
\nonumber
\ee
\be
=\frac{V(q)}{i\beta\omega}\left[(A_{B\rho}(q,\omega)
C_{B\rho}(q,\omega )
-A_{\rho B}(q,\omega)
C_{\rho B}(q,\omega ))\right]
\nonumber
\ee
\be
=\frac{1}{i\beta\omega}\left[D_{B\rho}(q,\omega )
-D_{\rho B}(q,\omega )\right]
\ee
or
\be
i\beta \omega D_{\rho\rho}(q,\omega )=
D_{B\rho}(q,\omega )
-D_{\rho B}(q,\omega )
\label{eq:616}
\ee
and
\be
-i\beta \omega D^{*}_{\rho\rho}(q,\omega )=
D_{\rho B}(q,\omega )
-D_{B\rho }(q,\omega )
\label{eq:617}
\ee
Together Eqs.(\ref{eq:616}) and (\ref{eq:617})
give
\be
D_{\rho\rho}(q,\omega )=
D^{*}_{\rho\rho}(q,\omega )
\ee
and
\be
D_{\alpha\beta}(q,\omega )=
D^{*}_{\beta\alpha}(q,\omega )
~~~.
\ee
Then
\be
\bar{G}_{\alpha\beta}(q,\omega )={\cal G}_{\alpha \mu}(q,\omega )
\sigma_{\mu\nu}(q)G_{\nu \beta}(q,\omega)
\ee
is a FDRMP, as is
\be
\tilde{G}_{\alpha\beta}(q,\omega )=\bar{G}_{\alpha \mu}(q,\omega )
\sigma_{\mu\nu}(q){\cal G}_{\nu \beta}(q,\omega)
~~~.
\ee
In operator notation
\be
\tilde{G}=
{\cal G}\sigma G \sigma {\cal G}
~~~.
\ee


\begin{thebibliography}{99}

\bibitem{BB} L. Berthier and G. Biroli, Rev. Mod. Phys. {\bf 83}, 587 (2011).

\bibitem{NDI}
S. P. Das and G. F. Mazenko, J. Stat. Phys. vol. {\bf 149}, issue 4, 643-657 (2012).
Referred to as NDI.

\bibitem{NDII}
S. P. Das and G. F. Mazenko, arXiv:1303.1627 (2013). (\textit{Submitted to J. Stat. Phys.})
Referred to as NDII

\bibitem{FTSPD} G. F. Mazenko, Phys. Rev. E {\bf 81}, 061102 (2010).
Referred to as FTSPD

\bibitem{SDENE} G. F. Mazenko, Phys. Rev. E {\bf 83}, 041125 (2011).
Referred to as SDENE.

\bibitem{FPD}
Fokker-Planck dynamics is Newtonian dynamics supplemented by viscous force
terms.

\bibitem{MMSI}G. F. Mazenko, D. D. McCowan, and P. Spyridis,
Phys. Rev. E {\bf 85}, 051105 (2012).
Referred to as MMS.

\bibitem{SM} P. Spyridis and G. F. Mazenko, arXiv:1302.2281 (2013). (\textit{Submitted to J. Stat. Phys.})
Referred to as SM.

\bibitem{MCT1}
W. Goetze  in
{\bf Liquids, Freezing and Glass Transition},
edited by Hansen J. P., Levesque D. and Zinn-Justin
J. (North-Holland, Amsterdam) 1991

\bibitem{MCT2}
S. Das, Rev. Mod. Phys. {\bf 76}, 785 (2004).

\bibitem{MSR}
MSR actions and conjugate fields  are discussed in Appendix A in FTSPD.

\bibitem{MMS2}
G. F. Mazenko, D. D. McCowan, and P. Spyridis,
Unpublished.

\bibitem{VC} A careful analysis in ND2 of the vertex
corrections at second order
shows that vertex corrections,
because of the FDR, are simpler than at first sight.
The role of vertex corrections at higher-order will be addressed elsewhere.


\bibitem{RM} One finds siimilar static equations from metastable states
using replica methods.
G. Parisi and Tammasso Rizzo, Phys. Rev. E {\bf 87} 012101 (2013).
Hugo Joachim and Francesco Zamponi, arXiv:1211.3468 (2012).

\bibitem{vS} The von Schweidler contribution shows a time dependence
  $\phi_{\nu}(t)=-t^{b}$
and $\lambda =\frac{\Gamma^{2}(1+b)}{\Gamma (1+2b)}$.
For more details see Ref.(\onlinecite{SM}).

\end{thebibliography}
\end{document}